\documentclass[11pt]{article}

\usepackage{amsfonts}
\usepackage{amsmath} 
\usepackage{amssymb}
\usepackage{enumerate}

\newcommand{\PROOF}{\begin{proof}}

\newcommand{\QED}{\end{proof}}

\newcommand{\ceil}[1]{ \left\lceil #1 \right\rceil }

\newcommand{\myset}[2]{ \left\{ #1 \left| #2 \right. \right\} }

\newcommand{\prefix}{\sqsubseteq}

\newcommand{\limn}{\lim\limits_{n\to\infty}}
\newcommand{\liminfn}{\liminf\limits_{n\to\infty}}
\newcommand{\limsupn}{\limsup\limits_{n\to\infty}}

\newcommand{\restr}{\upharpoonright}

\newcommand{\N}{\mathbb{N}}

\newcommand{\Q}{\mathbb{Q}}

\newcommand{\comp}{{\mathrm{comp}}}

\newcommand{\dimh}{\mathrm{dim}_\mathrm{H}}

\newcommand{\dimcomp}{\mathrm{dim}_\comp}
\newcommand{\cdim}{\mathrm{cdim}}

\renewcommand{\dim}{{\mathrm{dim}}}

\newcommand{\Dim}{{\mathrm{Dim}}}
\newcommand{\cDim}{{\mathrm{cDim}}}

\newcommand{\str}{{\mathrm{str}}}
\newcommand{\regSS}{S^\infty}
\newcommand{\strSS}{S^\infty_{\mathrm{str}}}

\newcommand{\C}{\mathbf{C}}
\newcommand{\bd}{{\mathbf{d}}}

\newcommand{\ALL}{\mathrm{ALL}}
\newcommand{\RAND}{\mathrm{RAND}}

\newcommand{\DIM}{\mathrm{DIM}}
\newcommand{\DIMstr}{\DIM_\str}

\newcommand{\DEC}{\mathrm{DEC}}

\newcommand{\calP}{{\cal P}}

\usepackage{amsthm}

\newtheorem{theorem}{Theorem}[section]
\newtheorem{corollary}[theorem]{Corollary}
\newtheorem{lemma}[theorem]{Lemma}
\newtheorem{proposition}[theorem]{Proposition}

\newenvironment{corollary_cite}[1]
{\begin{corollary}  {\rm (#1)}}
{\end{corollary}}

\newtheorem{question}[theorem]{Question}

\newtheorem*{definition}{Definition}

\theoremstyle{definition}

\newtheorem{example}[theorem]{Example}

\newtheorem*{example*}{Example}
\newtheorem*{examples*}{Examples}

\newtheorem*{ack}{Acknowledgment}

\theoremstyle{remark}

\numberwithin{equation}{section}
\numberwithin{figure}{section}

\usepackage{fullpage}
\usepackage{amsmath}
\usepackage{amssymb}
\usepackage{enumerate}

\def\bmath#1{\mbox{\boldmath$#1$}}
\newcommand{\bfSigma}{\bmath{\Sigma}}
\newcommand{\bfPi}{\bmath{\Pi}}

\newcommand{\cmp}[1]{\overline{#1}}
\newcommand{\dom}{{\rm dom}}
\newcommand{\concat}{\,\hat{}\,}

\newcommand{\Cant}{\mathbf{C}}

\newcommand{\A}{{\mathcal A}}
\newcommand{\B}{{\mathcal B}}
\renewcommand{\C}{{\mathcal C}}
\newcommand{\OO}{{\mathcal O}}
\newcommand{\U}{{\mathcal U}}
\newcommand{\V}{{\mathcal V}}
\newcommand{\X}{{\mathcal X}}
\newcommand{\Y}{{\mathcal Y}}

\newcommand{\RANDSch}{\RAND_{\mathrm{Schnorr}}}
\newcommand{\RANDcomp}{\RAND_\comp}

\title{\bf The Arithmetical Complexity of Dimension and Randomness}

\date{}

\author{
John M. Hitchcock
\thanks{
Department of Computer Science,
University of Wyoming,
Laramie, WY 82071, USA.
\texttt{jhitchco@cs.uwyo.edu}.
This author's research was supported in part by National Science Foundation
grant 9988483.}
\and
Jack H. Lutz
\thanks{ 
Department of Computer Science, Iowa
State University, Ames, IA 50011, USA. \texttt{lutz@cs.iastate.edu}.  
This author's research was supported in part by 
National Science
Foundation grants 9988483 and 0344187.} 
\and
Sebastiaan A. Terwijn\thanks{
Technische Universit\"at Wien,
Wiedner Hauptstrasse 8-10,
A-1040 Vienna, Austria. 
\texttt{terwijn@logic.at}.
Supported by the Austrian Research Fund (Lise Meitner grant M699-N05). 
Part of this author's research was done while
visiting the second author at Caltech in January 2002.}
}
 
\begin{document}

\maketitle
\begin{abstract} \noindent
Constructive dimension and constructive strong dimension are
effectivizations of the Hausdorff and packing dimensions,
respectively.  Each infinite binary sequence $A$ is assigned a
dimension $\dim(A) \in [0,1]$ and a strong dimension $\Dim(A) \in
[0,1]$.

Let $\DIM^\alpha$ and $\DIMstr^\alpha$ be the classes of all sequences
of dimension $\alpha$ and of strong dimension $\alpha$, respectively.
We show that $\DIM^0$ is properly $\Pi^0_2$, and that for all
$\Delta^0_2$-computable $\alpha \in (0,1]$, $\DIM^\alpha$ is properly
$\Pi^0_3$.

To classify the strong dimension classes, we use a more powerful
effective Borel hierarchy where a co-enumerable predicate is used rather
than a enumerable predicate in the definition of the $\Sigma^0_1$ level.
For all $\Delta^0_2$-computable $\alpha \in [0,1)$, we show that
$\DIMstr^\alpha$ is properly in the $\Pi^0_3$ level of this hierarchy.
We show that $\DIMstr^1$ is properly in the $\Pi^0_2$ level of this
hierarchy.

We also prove that the class of Schnorr random sequences and the class
of computably random sequences are properly $\Pi^0_3$.

\end{abstract}


\section{Introduction}


Hausdorff dimension -- the most extensively studied fractal dimension
-- has recently been effectivized at several levels of complexity,
yielding applications to a variety of topics in theoretical computer
science, including data compression, polynomial-time degrees,
approximate optimization, feasible prediction, circuit-size
complexity, Kolmogorov complexity, and randomness
\cite{Lutz:DCC,Lutz:DISS,Dai:FSD,AmMeReSt01,Hitchcock:MEHA,Fortnow:PD,Hitchcock:FDLLU,Mayordomo:KCCCHD}.
The most fundamental of these effectivizations is {\em constructive
  dimension}, which is closely related to Kolmogorov complexity and
algorithmic randomness.  Every subset $\X$ of $\Cant$, the {\em
Cantor space} of all infinite binary sequences, is assigned 
a constructive dimension $\cdim(\X) \in [0,1]$.  Informally, this
dimension is determined by the maximum rate of growth that a lower
semicomputable martingale can achieve on all sequences in $\X$.

Just as Martin-L\"of \cite{MartinLof66} used constructive measure to
define the randomness of individual sequences, Lutz \cite{Lutz:DISS}
used constructive dimension to define the dimensions of individual
sequences.  Each sequence $A \in \Cant$ is assigned a {\em dimension}
$\dim(A) \in [0,1]$ by $\dim(A) = \cdim(\{A\})$.  Every Martin-L\"of
random sequence has dimension 1, but there are nonrandom sequences
with dimension 1.  For every real number $\alpha \in [0,1]$, there is
a sequence with dimension $\alpha$.

It is useful to understand the arithmetical complexity of a
class of sequences.  For example, knowing that $\RAND$, the class of
Martin-L\"of random sequences, is a $\Sigma^0_2$-class allows the
application of Kreisel's Basis Lemma \cite{Krei50,Odif89} to give a
short proof \cite{vanL87a} that
\begin{equation}\label{eq:RAND_Delta}
\RAND \cap \Delta^0_2 \not= \emptyset.
\end{equation}
For any $\alpha \in [0,1]$, let
$$\DIM^\alpha = \{A \in \Cant \mid \dim(A) = \alpha\}.$$ 
Lutz \cite{Lutz:DISS} showed that 
\begin{equation}\label{eq:DIM_Delta}
\DIM^\alpha \cap \Delta^0_2 \not= \emptyset
\end{equation}
for any $\Delta^0_2$-computable $\alpha \in [0,1]$.  As these dimension classes
do not appear to be $\Sigma^0_2$, Lutz was unable to
apply the Basis Lemma to them, so he used different techniques to
prove \eqref{eq:DIM_Delta}.

We investigate the complexities of these dimension classes in terms of
the arithmetical hierarchy of subsets of $\Cant$.  
We show that $\DIM^0$ is properly $\Pi^0_2$, and for all
$\Delta^0_2$-computable $\alpha \in (0,1]$ we show that $\DIM^\alpha$
is properly $\Pi^0_3$.  
Therefore, the proof for \eqref{eq:RAND_Delta}
using Kreisel's Basis Lemma cannot be used directly to establish
\eqref{eq:DIM_Delta}. (See however the comments made after Corollary~\ref{proper}.)


More recently, {packing dimension}, another important fractal
dimension, has also been effectivized by Athreya, Hitchcock, Lutz, and
Mayordomo \cite{Athreya:ESDAICC}.  At the constructive level, this is
used in an analogous way to define the {\em strong dimension} $\Dim(A)
\in [0,1]$ for every sequence $A$.  For any $\alpha \in [0,1]$, let
$$\DIMstr^\alpha = \{A \in \Cant \mid \Dim(A) = \alpha\}.$$ To
classify these strong dimension classes, we use a more
powerful effective Borel hierarchy where a co-enumerable predicate is
used rather than a enumerable predicate in the definition of the
$\Sigma^0_1$ level.  We show that $\DIMstr^1$ is properly in the
$\Pi^0_2$ level of this stronger hierarchy.  For all
$\Delta^0_2$-computable $\alpha \in [0,1)$, we show that
$\DIMstr^\alpha$ is properly in the $\Pi^0_3$ level of this hierarchy.

Our techniques for classifying the dimension and strong dimension
classes include Baire category, Wadge reductions, and Kolmogorov
complexity. In Section \ref{adhoc} we point out that 
ad hoc methods are sometimes necessary. 

Section \ref{sec:background} gives an overview of the randomness and
dimension notions used in this paper.  In Section \ref{sec:borel} we
introduce the stronger effective Borel hierarchy that we use for the
strong dimension classes.  Section \ref{sec:dimclass} presents the
classification of $\DIM^\alpha$ and $\DIMstr^\alpha$.  

We conclude the paper with Section \ref{sec:randclass} on 
effective randomness classes.
We restate a result of Schnorr \cite{Schn71b} concerning computable
null sets of exponential order in terms of {\em computable dimension}
and point out a relationship with {\em Church randomness}.  We
prove that the class of {\em Schnorr random} sequences and that the
class of {\em computably random} sequences are properly $\Pi^0_3$.

\section{Background on Randomness and Dimension} \label{sec:background}

This section provides an overview of the notions of randomness and
dimension used in this paper.  We write $\{0,1\}^*$ for the set of all
finite binary {\em strings} and $\Cant$ for the {\em Cantor space} of
all infinite binary {\em sequences}.  In the standard way, a sequence
$A \in \Cant$ can be identified with the subset of $\{0,1\}^*$ or $\N$
for which it is the characteristic sequence, or with a real number in
the unit interval.  The length of a string $w \in \{0,1\}^*$ is $|w|$.
The string consisting of the first $n$ bits of $x \in \{0,1\}^* \cup \Cant$ is
denoted by $x\restr n$.  We write $w \prefix x$ if $w$ is a prefix of $A$.

\subsection{Martin-L\"of Randomness}

Martin-L\"of \cite{MartinLof66} introduced the notion of a {\em
constructive null set\/}.  A set is constructively null if it can be
covered by a uniform sequence of c.e. open sets that are shrinking in
size.
That is, $\A \subseteq \Cant$ is {constructive null} if
$\A\subseteq \bigcap_i \U_i$, where $\{\U_i\}_{i\in\N}$ is uniformly
c.e.\ such that $\mu(\U_i)\leq 2^{-i}$.
The sequence $\{\U_i\}_{i\in\N}$ is called a {\em Martin-L\"of test}.
An individual sequence $A \in \Cant$ is {\em Martin-L\"of random\/} if
$\{A\}$ is not constructively null.
The Martin-L\"of random sequences play an important role in
algorithmic information theory, see e.g.\ Li and Vit\'anyi
\cite{LiVi97}.

Schnorr \cite{Schn71b}, following Ville \cite{Vill39}, characterized
constructive null sets in terms of martingales.  A function
$d:\{0,1\}^*\rightarrow [0,\infty)$ is a {\em martingale\/} if for
every $w\in \{0,1\}^*$, $d$ satisfies the averaging condition
\begin{equation*} \label{mart}
2d(w) = d(w0) + d(w1),
\end{equation*}
and $d$ is a {\em supermartingale\/} if it satisfies
\begin{equation*} \label{supmart}
2d(w) \geq d(w0) + d(w1). 
\end{equation*}
The {\em success set of $d$} is 
$$\regSS[d] = \myset{ A \in \Cant }{ \limsupn d(A\restr n) =
  \infty},$$ i.e., it is the set of all sequences on which $d$ has
  unbounded value.  We say that $d$ succeeds on a class $\A\subseteq
  \Cant$ if $\A \subseteq \regSS[d]$.

Ville \cite{Vill39} proved that a set $\A \subseteq \Cant$ has
Lebesgue measure 0 if and only if there is a martingale $d$ that
succeeds on $\A$.  Schnorr \cite{Schn71b} showed that $\A$ is
constructively null if and only if $d$ can be chosen to be lower
semicomputable, that is, if $d$ can be computably approximated from
below.  We call such a $d$ {\em constructive}.

Martin-L\"of \cite{MartinLof66} proved that there is a universal
constructive null set. That is, he proved that there is a Martin-L\"of
test $\{\U_i\}_i$ such that for every other test $\{\V_i\}$ it holds
that $\bigcap_i \V_i\subseteq \bigcap_i \U_i$.  By Schnorr's analysis
this implies that there is also a universal constructive
supermartingale $\bd$.  That is, for any constructive supermartingale
$d'$ there is a $c > 0$ such that $\bd(w) \geq c d'(w)$ for all $w \in
\{0,1\}^*$.  We will use this universal supermartingale in
section~\ref{sec:dimclass}.  We denote the complement of
$\regSS[\bd]$ by $\RAND$, so that $\RAND$ consists of all the
Martin-L\"of random sequences.

\subsection{Schnorr Randomness} \label{Snull}

Schnorr \cite{Schn71b} criticized the notion of constructive null for
an actual lack of constructiveness, and introduced the more
constructive notion of a {\em Schnorr null set}, which is defined by
requiring that the measure of the levels $\U_i$ in a Martin-L\"of test
be computably approximable to within any given precision.  It is easy
to see that this is equivalent to the following: $\A$ is Schnorr null
if $\A\subseteq \bigcap_i \U_i$, where $\{\U_i\}_{i\in\N}$ is
uniformly c.e.\ such that $\mu(\U_i)=2^{-i}$.  The sequence
$\{\U_i\}_{i\in\N}$ is called a {\em Schnorr test}.

Following Schnorr \cite{Schn71b}, we call an unbounded nondecreasing
function $h:\{0,1\}^*\rightarrow\{0,1\}^*$ an {\em order}.  (N.B. An
``Ordnungsfunktion'' in Schnorr's terminology is always computable,
whereas we prefer to leave the complexity of orders unspecified in
general.)  For any order $h$ and martingale $d$, we define the {\em order
$h$ success set of $d$} as 
$$S^h[d]=\myset{A\in\Cant}{\limsup_{n\rightarrow\infty} \frac{d(A\restr
n)}{h(n)}\geq 1}.$$ Schnorr pointed out that the rate of success of a
constructive martingale $d$ can be so slow that it cannot be
computably detected.  Thus rather than working with constructive null
sets of the form $\regSS[d]$ with $d$ constructive, he worked with
null sets of the form $S^h[d]$, where both $d$ and $h$ are computable.
He proved that a set $\A$ is Schnorr null if and only if it is
included in a null set of the form $S^h[d]$, with $d$ and $h$
computable.

A sequence $A \in \Cant$ is {\em Schnorr random\/} if $\{A\}$ is not
Schnorr null.  This is related the notion of computable randomness.  A
sequence $A$ is {\em computably random\/} if for every computable
martingale $d$, $A \not\in \regSS[d]$.

We write $\RANDSch$ for the class of all Schnorr random sequences and
$\RANDcomp$ for the class of all computably random sequences.
By definition we have that
$$\RAND \subseteq \RANDcomp \subseteq \RANDSch.$$ The first inclusion
was proved strict by Schnorr \cite{Schn71b} and the second inclusion
was proved strict by Wang \cite{Wang96b}.

\subsection{Constructive Dimension}

Hausdorff \cite{Haus19} introduced the concept of null covers that
``succeed exponentially fast'' to define what is now commonly called
Hausdorff dimension, the most widely used dimension in fractal
geometry.  Basically, this notion allows one to discern structure in
classes of measure zero, and to calibrate them.  As for constructive
measure, already Schnorr (see Theorem~\ref{exponential}) drew special
attention to null sets of ``exponential order'', although he did not
make an explicit connection to Hausdorff dimension.

Lutz \cite{Lutz:DCC,Lutz:DISS} gave a characterization of Hausdorff
dimension in terms of {\em gales}, which are a generalization of
martingales.  Let $s\in [0,\infty)$. An {\em $s$-gale\/} is a function
$d:\{0,1\}^*\rightarrow [0,\infty)$ that satisfies the averaging
condition
\begin{equation} \label{gale}
2^sd(w) = d(w0) + d(w1)
\end{equation}
for every $w\in \{0,1\}^*$.  Similarly, $d$ is an {\em
$s$-supergale\/} if (\ref{gale}) holds with $\geq$ instead of
equality.  The success set $\regSS[d]$ is defined exactly as was done
for martingales above.  Lutz showed that for any class $\A \subseteq
\Cant$, the Hausdorff dimension of $\A$ is 
\begin{equation}\label{eq:dimh}
\dimh(\A) = \inf \left\{ s \left| \begin{array}{l}\textrm{there
exists an $s$-gale}\\\textrm{$d$ for which $\A \subseteq
S^\infty[d]$}\end{array} \right.\right\}.
\end{equation}
Lutz \cite{Lutz:DISS} effectivized this characterization to define the
{constructive dimensions} of sets and sequences.  An $s$-(super)gale
is called {\em constructive} if it is lower semicomputable.  The
{\em constructive dimension} of a class $\A \subseteq \Cant$ is
\begin{equation}\label{eq:cdim}
\cdim(\A) = \inf \left\{ s \left| \begin{array}{l}\textrm{there
exists a constructive $s$-gale}\\\textrm{$d$ for which $\A \subseteq
S^\infty[d]$}\end{array} \right.\right\}
\end{equation}
 and the constructive
dimension of an individual sequence $A \in \Cant$ is $$\dim(A) =
\cdim(\{A\}).$$ (Supergales can be equivalently used in place of gales
in both \eqref{eq:dimh} and
\eqref{eq:cdim} \cite{Lutz:DCC,Hitchcock:GSCD,Fenn02}.)  

Constructive dimension has some remarkable properties.  For example,
Lutz \cite{Lutz:DISS} showed that for any class $\A$,
\begin{equation}\label{eq:dim_pointwise}
\cdim(\A) = \sup_{A \in \A} \dim(A).
\end{equation}  Also, Mayordomo \cite{Mayordomo:KCCCHD}
established a strong connection with {\em Kolmogorov complexity}: for any
$A \in \Cant$,
\begin{equation}\label{eq:K_dim}
\dim(A) = \liminfn \frac{K(A \restr n)}{n},
\end{equation}
where $K(A \restr n)$ is the size of the smallest program that causes
a fixed universal self-delimiting Turing machine to output the first
$n$ bits of $A$.  (For comments on the relation of this result to
earlier results, see the report \cite{Stai03} by Staiger and section 6
of \cite{Lutz:DISS}.  For more details on Kolmogorov complexity, we
refer to \cite{LiVi97}.)

One can also characterize constructive dimension using the Schnorr
null sets (see Section~\ref{Snull}) of exponential order. The
following proposition was observed by several authors, including those
of \cite{AmMeReSt01,Terw03}.

\begin{proposition} \label{equiv}
Let $\bd$ be the universal constructive supermartingale.  For any $\A
\subseteq \Cant$,
$$
\cdim(\A) =
\inf\{s\in\Q:
\A\subseteq S_{2^{(1-s)n}}[\bd] \,) \}.
$$
\end{proposition}

\subsection{Constructive Strong Dimension}

More recently, Athreya, Hitchcock, Lutz, and Mayordomo
\cite{Athreya:ESDAICC} also characterized {\em packing dimension},
another important fractal dimension, in terms of gales.  For this, the
notion of {\em strong success} of an $s$-gale $d$ was introduced.  The
{\em strong success set of $d$} is
$$\strSS[d] = \myset{A \in \Cant }{ \liminfn d(A \restr n) = \infty}.$$
Analogously to what was done for Hausdorff dimension, packing
dimension can be characterized using strong success sets of gales.
Effectivizing this in the same way leads to the definition of the {\em
constructive strong dimension} of a class $\A \subseteq \Cant$ as
\begin{equation*}\label{eq:cDim}
\cDim(\A) = \inf \left\{ s \left| \begin{array}{l}\textrm{there
exists a constructive $s$-gale}\\\textrm{$d$ for which $\A \subseteq
\strSS[d]$}\end{array} \right.\right\}.
\end{equation*}
The constructive strong dimension of a sequence $A \in \Cant$ is
\begin{equation*}\label{eq:Dim}
\Dim(A) = \cDim(\{A\}).
\end{equation*}
A pointwise stability property analogous to \eqref{eq:dim_pointwise}
also holds for strong dimension, as well as a Kolmogorov
complexity characterization \cite{Athreya:ESDAICC}:
\begin{equation}\label{eq:K_Dim}
\Dim(A) = \limsupn \frac{K(A\restr n)}{n}
\end{equation}
for any $A \in \Cant$.

\section{Borel Hierarchies}\label{sec:borel}


$\bmath{\Sigma}^0_n$ and $\bmath{\Pi}^0_n$ denote the levels of the
Borel hierarchy for subsets of Cantor space.  The levels of the
arithmetical hierarchy 
(the corresponding effective hierarchy for sets of reals) are
denoted by $\Sigma^0_n$ and $\Pi^0_n$.

We will also make use of the following more general hierarchy
definition.

\begin{definition}  Let $\calP$ be a class of predicates, let 
$n \geq 1$, and let $\X \subseteq \Cant$.
\begin{itemize}
\item $\X \in \Sigma^0_n[\calP]$ if for some predicate $P \in \calP$,
$$A \in \X \iff (\exists k_n) (\forall k_{n-1}) \cdots (Q k_1)
  P(k_n,\ldots,k_2,A\restr k_1),$$
where $Q = \exists$ if $n$ is odd  and  $Q = \forall$ if $n$ is even.
\item $\X \in \Pi^0_n[\calP]$ if for some predicate $P \in \calP$,
$$A \in \X \iff (\forall k_n) (\exists k_{n-1}) \cdots (Q k_1)
  P(k_n,\ldots,k_2,A\restr k_1),$$
where $Q = \forall$ if $n$ is odd  and  $Q = \exists$ if $n$ is even.
\end{itemize}
\end{definition}

If we take $\calP$ to be $\Delta^0_1$ (decidable), then the
above definition is equivalent to the standard arithmetical hierarchy 
of reals, that is 
$$\Sigma^0_n = \Sigma^0_n[\Delta^0_1]$$
and
$$\Pi^0_n = \Pi^0_n[\Delta^0_1]$$
hold for all $n$.
Also, if $\ALL$ is the class of all predicates, then we obtain the
classical Borel hierarchy:
         $${\bfSigma^0_n} = \Sigma^0_n[\ALL]$$
and
             $${\bfPi^0_n} = \Pi^0_n[\ALL].$$

In this paper, we will also be interested in the cases where $\calP$
is $\Sigma^0_1$ (computably enumerable) or $\Pi^0_1$ (co-c.e.).  
In some cases, the classes in the generalized hierarchy
using these sets of predicates are no different than the standard
arithmetical hierarchy classes.  If $n$ is odd, then $\Sigma^0_n =
\Sigma^0_n[\Sigma^0_1]$ as the existential quantifier in the
$\Sigma^0_1$ predicate can be absorbed into the last quantifier in the
definition of $\Sigma^0_n[\Delta^0_1] = \Sigma^0_n$.  Analogously,
$\Pi^0_n = \Pi^0_n[\Pi^0_1]$ for odd $n$, and for even $n$ we have
$\Sigma^0_n = \Sigma^0_n[\Pi^0_1]$ and $\Pi^0_n =
\Pi^0_n[\Sigma^0_1].$ On the other hand, using the complementary set
of predicates defines an effective hierarchy that is distinct from and
interleaved with the arithmetical hierarchy.

\begin{proposition}\label{prop:borel}
\begin{enumerate}
\item If $n$ is odd, then 
$$\Sigma^0_n \subsetneq \Sigma^0_n[\Pi^0_1]
\subsetneq \Sigma^0_{n+1}$$
and
$$\Pi^0_n \subsetneq \Pi^0_n[\Sigma^0_1] \subsetneq
\Pi^0_{n+1}.$$
\item If $n$ is even, then 
$$\Sigma^0_n \subsetneq \Sigma^0_n[\Sigma^0_1]
\subsetneq \Sigma^0_{n+1}$$
and
$$\Pi^0_n \subsetneq \Pi^0_n[\Pi^0_1] \subsetneq
\Pi^0_{n+1}.$$
\end{enumerate}
\end{proposition}

\begin{proof}
We only show $\Sigma^0_n \subsetneq \Sigma^0_n[\Pi^0_1] \subsetneq
\Sigma^0_{n+1}$ for odd $n$; the arguments for the other statements
are analogous.

The inclusion $\Sigma^0_n \subseteq \Sigma^0_n[\Pi^0_1]$ is obvious. 
To show that it is proper, let $P$ be a predicate
that is complete for the class of $\Pi^0_n$ predicates.  Then there is
a decidable predicate $R$ such that
$$P(n) \iff (\forall k_n)(\exists k_{n-1})\cdots(\forall k_1)
R(n,k_n,\cdots,k_1).$$
Define $\X \subseteq \Cant$ as
$$\X = \bigcup_{n\in P} 0^n 1\Cant.$$
Then $\X \in \Sigma^0_n[\Pi^0_1]$ as we have 
\begin{eqnarray*}
S \in \X &\iff& (\exists n) P(n) \textrm{ and } 0^n1 \prefix S \\
&\iff& 
(\exists n)(\forall k_n)(\exists k_{n-1})\cdots(\forall k_1)
R(n,k_n,\cdots,k_1) \textrm{ and }0^n1 \prefix S \\
&\iff& 
(\exists n)(\forall k_n)(\exists k_{n-1})\cdots(\exists k_2)
T(n,k_n,\cdots,k_3,S\restr k_2),
\end{eqnarray*}
where $T$ is the $\Pi^0_1$ predicate defined by
$$T(n,k_n,\cdots,k_3,w) \iff (\forall k_1)
R(n,k_n,\cdots,k_3,|w|,k_1)\textrm{ and }0^n1 \prefix w.$$
Now suppose that $\X \in \Sigma^0_n$.  Then for some decidable
predicate $U$, 
$$S \in X \iff 
(\exists k_n)(\forall k_{n-1})\cdots(\exists k_1)
U(k_n,\cdots,k_2,S\restr k_1).$$
We then have 
\begin{eqnarray*}
n \in P &\iff& 0^n1\Cant \subseteq \X \\
&\iff& 0^n 1 0^\infty \in \X \\
&\iff& 
(\exists k_n)(\forall k_{n-1})\cdots(\exists k_1)
U(k_n,\cdots,k_2,0^n10^\infty\restr k_1),
\end{eqnarray*}
so $P$ is a $\Sigma^0_n$ predicate, which contradicts its
$\Pi^0_n$-completeness.  Therefore $\X \not\in \Sigma^0_n$ and we have
established $\Sigma^0_n \subsetneq \Sigma^0_n[\Pi^0_1]$.  

The inclusion $\Sigma^0_n[\Pi^0_1] \subseteq \Sigma^0_{n+1}$ is
immediate from the definitions using $\Sigma^0_{n+1} =
\Sigma^0_{n+1}[\Delta^0_1]$.  That it is proper follows from the facts
$\Sigma^0_{n+1} - {\bfSigma^0_n} \not= \emptyset$ and
$\Sigma^0_n[\Pi^0_1] \subseteq {\bfSigma^0_n}$.
\end{proof}

The next proposition shows that there are no unexpected 
inclusions: 

\begin{proposition}
\begin{enumerate}
\item If $n$ is odd, then
$$\Sigma^0_n \not\subseteq \Pi^0_n[\Sigma^0_1] 
\not\subseteq \Sigma^0_{n+1} $$
and
$$\Pi^0_n \not\subseteq \Sigma^0_n[\Pi^0_1]
\not\subseteq \Pi^0_{n+1}.$$

\item If $n$ is even, then
$$\Sigma^0_n \not\subseteq \Pi^0_n[\Pi^0_1]
\not\subseteq  \Sigma^0_{n+1}$$
and
$$\Pi^0_n \not\subseteq \Sigma^0_n[\Sigma^0_1]
\not\subseteq \Pi^0_{n+1}.$$
\end{enumerate}
\end{proposition}
\begin{proof}
The noninclusions on the left side all follow from 
Borel considerations. E.g.\ for the noninclusion 
$\Pi^0_n \not\subseteq \Sigma^0_n[\Pi^0_1]$ take 
any $\Pi^0_n$-class that is not in $\bfSigma^0_n$. 

The noninclusions on the right side can all be proved 
by direct diagonalization. As an example we prove 
that
$\Sigma^0_1[\Pi^0_1]\not\subseteq \Pi^0_2$. 
The proof easily generalizes to the higher levels. 
The proof is a fairly straightforward diagonalization against 
all possible $\Pi^0_2$-definitions, although the details 
are a bit cumbersome. Let $R_i$, $i\in\omega$, be a 
computable list of all partial computable predicates. 
We define a class 
$\X \in \Sigma^0_1[\Pi^0_1]$ such that for all $i$ there is 
$X \sqsupset 0^i1$ such that 
\begin{equation} \label{counterexample}
X \in  (\X\setminus\Y) \cup (\Y\setminus\X),
\end{equation} 
where $\Y = \big\{ X : \forall n \exists m \; R_i(n,X\restr m) \big\}$.
So the definition of $\X$ in the interval above the string $0^i1$
will make sure that $\X$ is not $\Pi^0_2$-defined by $R_i$. 

For the definition of $\X$ we will need a uniform sequence of
$\Pi^0_1$-sets of strings $P^i_n$.  We start by defining $\dom(P^i_n)$
for each $i$ and $n$ such that
\begin{itemize}
\item 
$\dom(P^i_n)$ is a computable subset of
$\big\{ \sigma1: \sigma\sqsupseteq 0^i1 \big\}$,
\item $\dom(P^i_n)$ is dense above the string $0^i1$, i.e.\ for all 
$\tau \sqsupseteq 0^i1$ there is $\sigma\sqsupseteq \tau$ with $\sigma\in \dom(P^i_n)$,
\item $\dom(P^i_n) \cap \dom(P^i_m) = \emptyset$ if $n\neq m$.
\end{itemize}
Then we define $P^i_n$ by 
\begin{itemize}
\item for every $\sigma1 \in \dom(P^i_n)$,  
\begin{equation} \label{defP}
\sigma1\not\in P^i_n \Longleftrightarrow 
(\exists \tau \sqsupseteq \sigma1)( \exists m) \; [ \:  \tau \mbox{ is of the form 
$\sigma10^k$} \wedge R_i(n, \tau \restr m) \: ].
\end{equation}
\end{itemize}
It is easy to see that such a uniform sequence of $P^i_n$'s exists. 
Now $\X$ is defined as 
$$
\X = \big\{X : (\exists i)(\exists n)( \exists \sigma \in P^i_n) \; 
[\, \sigma \sqsubset X \, ]  \big\}.
$$
The idea is that to show that $R_i$ does not give a $\Pi^0_2$-definition of $\X$, 
we challenge it by choosing a string $\sigma 1 \in P^i_n$ and extend it by $0$'s. 
Now if $R_i$ responds by providing us with an extension $\tau$ as in (\ref{defP}),
we take this $\tau$ as an initial segment of our set $X$, 
which means that there is an $m$ such that $R_i(n, X\restr m)$, so that the 
condition $\forall n \exists m \; R_i(n,X\restr m)$ is verified for $n$. 
But by definition of $P^i_n$, as soon as the witness $m$ is found, the 
string $\sigma1$ falls out of the set $P^i_n$, so we have mananged to keep 
$X$ outside of $\X$ while at the same time obtaining a piece of evidence that 
$\forall n \exists m \; R_i(n,X\restr m)$. If on the other hand $R_i$ does not 
respond this means that $\sigma1\in P^i_n$, so $X$ will be in $\X$, but {\em no\/} 
extension $Y$ of $\sigma1$ will satisfy $\forall n \exists m \; R_i(n,Y\restr m)$.
So in both cases $X$ is a counterexample showing that $R_i$ does not 
define $\X$. 

We now give the formal construction. Fix $i$. We construct $X$ as in (\ref{counterexample}) 
by a finite extension construction. Let $X_0 = 0^i1$. 
At stage $s$ of the construction we are given $X_s$, no initial segment of which 
is in any $P^i_n$, and such that 
$(\forall n \leq s)(\exists m \leq |X_s|) \; [ \: R_i(n, X_s \restr m) \: ]$.
Choose $\sigma1\in \dom(P^i_{s+1})$ such that $\sigma1\sqsupset X_s$. 

Case I. There exists a $\tau$ as in (\ref{defP}), with $n= s+1$. Define $X_{s+1}= \tau$
and go to the next stage of the construction. 

Case II. There does not exist such a $\tau$. Then define $X = X_s\concat 0^\omega$ 
and end the construction. 

To verify that $X = \bigcup_s X_s$ thus constructed satisfies (\ref{counterexample}), 
note that if there is a stage where Case II obtains, then the string $\sigma1$ 
chosen at that stage is in $P^i_s$ and proves that $X\in\X$, whereas no 
extension $Y$ of $\sigma1$ satisfies $\exists m \; R_i(n,Y\restr m)$. 
So in this case we are done. If on the other hand at every stage of the 
construction Case I obtains, then for every $\sigma1$ chosen at any stage $s$ 
we have $\sigma1\not\in P^i_s$, and hence $X \not\in \X$ since apart from 
the $1$'s in $\sigma1$ the string $X$ only contains $0$'s, and 
$\dom(P^i_n) \subseteq \big\{ \sigma1: \sigma\sqsupseteq 0^i1 \big\}$.
But also at every stage $s$ a new witness $m$ is found such that 
$R_i(s, X\restr m)$, hence $\forall s \exists m \; R_i(s,X\restr m)$, 
and again $X$ satisfies (\ref{counterexample}). 
\end{proof}

Intuitively, the classes $\Sigma^0_1[\Pi^0_1],\ \Pi^0_1[\Sigma^0_1],\
\Sigma^0_2[\Sigma^0_1],\ \Pi^0_2[\Pi^0_1],\ldots\ $ are slightly more
powerful than their respective counterparts in the arithmetical
hierarchy because they use one additional quantifier that is limited
to the predicate.  We now give a simple example of a class that is
best classified in this hierarchy: the class of 1-generic sequences.

\begin{proposition}  \label{generic}
The class of all 1-generic
sequences is $\Pi^0_2[\Pi^0_1]$ but not $\Sigma^0_3$. 
It is also not $\bmath{\Sigma}^0_2$. 
\end{proposition}
\begin{proof}
Recall that a sequence $X \in \Cant$ is 1-generic  
(see e.g.\ Jockusch \cite{Jock80}) 
if
$$
(\forall e)(\exists\sigma\sqsubset X) \big [ \,
\{e\}^\sigma(e)\downarrow  \;\vee \;
(\forall \tau\sqsupset\sigma) [ \{e\}^\tau(e) \uparrow ] \,  \big ]
$$
From this definition it is immediate that the class ${\cal G} = \{X
\mid X \textrm{ is 1-generic}\}$ is in
$\Pi^0_2[\Pi^0_1]$. 
To show that ${\cal G}$ is not $\Sigma^0_3$, suppose that it is.
Then there is a uniform
sequence of $\Sigma^0_1$-classes $\OO_{n,m}$ such that
${\cal G}=\bigcup_n\bigcap_m\OO_{n,m}$. Without loss of
generality $\OO_{n,m}\supseteq \OO_{n,m+1}$ for all $n$,$m$.
Now ${\cal G}$ is comeager, so there is $n$ such that 
$\bigcap_m \OO_{n,m}$ is not nowhere dense, hence dense in
some interval $C_\sigma$. Then every $\OO_{n,m}$, $m\in\N$,
is dense in $C_\sigma$. Now it is easy to construct, using a 
computable finite extension construction, a computable 
sequence (starting with $\sigma$) in $\bigcap_m \OO_{n,m}$, contradicting
that 1-generic sets are noncomputable.

%
That the 1-generic sets are not $\mbox{\boldmath$\Sigma^0_2$}$ follows 
quickly from Lemma~\ref{le:category} below, noting again that the 1-generic 
sets are a comeager class. 
\end{proof}

Staiger has pointed out to us that the class $\Pi^0_1[\Sigma^0_1]$
already occured under a different guise in \cite{Staiger00} where it
was called $\mathfrak P$, and several presentations were proven to be
equivalent to it.  The following definitions are contained in
\cite{Staiger93}.  Let $W$ be any set of initial segments. Define
\begin{eqnarray*}
\lim W & = & \{A\in 2^\omega : \forall \sigma \sqsubset A (\sigma\in W) \}, \\
W^\sigma & = & \{A\in 2^\omega : \forall^\infty \sigma \sqsubset A (\sigma\in W)\}. 
\end{eqnarray*}
Staiger proved that the classes in $\Pi^0_1[\Sigma^0_1]$ are those of 
the form $\lim W$, for $W \in \Sigma^0_1$, and 
the classes in $\Sigma^0_2[\Sigma^0_1]$ are those of
the form $W^\sigma$, for $W \in \Sigma^0_1$.

\section{Classification of $\DIM^\alpha$ and $\DIMstr^\alpha$}
\label{sec:dimclass}

In this section we investigate the arithmetical complexity of the
following dimension and strong dimension classes.  
\begin{eqnarray*}
\DIM^\alpha &=& \{A \in \Cant \mid \dim(A) = \alpha\}\\
\DIM^{\leq\alpha} &=& \{A \in \Cant \mid \dim(A) \leq \alpha\}\\ 
\DIM^{\geq\alpha} &=& \{A \in \Cant \mid \dim(A) \geq \alpha\}\\ 
\DIMstr^\alpha &=& \{A \in \Cant \mid \Dim(A) = \alpha\}\\
\DIMstr^{\leq\alpha} &=& \{A \in \Cant \mid \Dim(A) \leq \alpha\}\\ 
\DIMstr^{\geq\alpha} &=& \{A \in \Cant \mid \Dim(A) \geq \alpha\}
\end{eqnarray*}

Let $\alpha \in [0,1]$ be $\Delta^0_2$-computable.  For any such
$\alpha$, it is well known that there is a computable function
$\hat{\alpha} : \N \to \Q$ such that $\limn \hat{\alpha}(n) = \alpha$.
Using \eqref{eq:K_dim}, we have
\begin{eqnarray*}
\dim(X) \leq \alpha &\iff& \liminfn \frac{K(X\restr n)}{n} \leq \alpha
\\
&\iff& (\forall k) (\forall N) (\exists n\geq N) K(X\restr n) < (\hat{\alpha}(n)+1/k) n,
\end{eqnarray*}
so $\DIM^{\leq \alpha}$ is a $\Pi^0_2$-class.  Also,
\begin{eqnarray*}
\dim(X) \geq \alpha &\iff& 
\liminfn \frac{K(X\restr n)}{n} \geq \alpha \\
&\iff& (\forall k) (\exists N) (\forall n \geq N) K(X\restr n) > (\hat{\alpha}(N)-1/k) n,
\end{eqnarray*}
so $\DIM^{\geq \alpha}$ is a $\Pi^0_3$-class.  Therefore we have the
following.

\begin{proposition}\label{prop:dim_member}
\begin{enumerate}
\item The class $\DIM^0$ is $\Pi^0_2$.
\item For all $\Delta^0_2$-computable $\alpha \in (0,1]$, $\DIM^\alpha$ is a
  $\Pi^0_3$-class. 
\item For arbitrary $\alpha \in (0,1]$, $\DIM^\alpha$ is a
$\bfPi^0_3$-class.
\end{enumerate}
\end{proposition}

The situation is slightly more complicated for strong dimension.  By
\eqref{eq:K_Dim}, we have
\begin{eqnarray*}
\Dim(X) \leq \alpha &\iff& \limsupn \frac{K(X\restr n)}{n} \leq \alpha
\\ 
&\iff& (\forall k) (\exists N) (\forall n\geq N)K(X\restr n) <
(\hat{\alpha}(N)+1/k) n\\ 
&\iff& (\forall k) (\exists N) (\forall
n\geq N) (\exists \langle \pi,t \rangle) |\pi| < (\hat{\alpha}(N)+1/k) n
\\&&\textrm{ and }U(\pi) = X\restr n\textrm{ in $\leq$ $t$ computation
steps},
\end{eqnarray*}
where $U$ is the fixed universal self-delimiting Turing machine used
to define $K$.  From this it is clear that $\DIMstr^{\leq \alpha} \in
\Pi^0_4$.  However, the ``$(\exists \langle \pi,t \rangle)$'' quantifier is local to
the defining predicate, so we have $\DIMstr^{\leq \alpha} \in
\bfPi^0_3$, and in fact, it is a $\Pi^0_3[\Sigma^0_1]$-class.  Also,
\begin{eqnarray*}
\Dim(X) \geq \alpha &\iff& 
\limsupn \frac{K(X\restr n)}{n} \geq \alpha \\
&\iff& (\forall k) (\forall N) (\exists n \geq N) K(X\restr n) > (\hat{\alpha}(n)-1/k) n,
\end{eqnarray*}
so $\DIMstr^{\geq \alpha}$ is a $\Pi^0_2[\Pi^0_1]$-class.  
This establishes the following analogue of Proposition
\ref{prop:dim_member}.
\begin{proposition}\label{prop:Dim_member}
\begin{enumerate}
\item The class $\DIMstr^1$ is
 $\Pi^0_2[\Pi^0_1]$.
\item For all $\Delta^0_2$-computable $\alpha \in [0,1)$, $\DIMstr^\alpha$
  is a $\Pi^0_3[\Sigma^0_1]$-class.  
\item
For arbitrary $\alpha \in [0,1)$, $\DIMstr^\alpha$ is a $\bfPi^0_3$-class.
\end{enumerate}
\end{proposition}

In the remainder of this section we 
prove that the classifications in Propositions
\ref{prop:dim_member} and \ref{prop:Dim_member} cannot be improved in
their respective hierarchies.

\subsection{Category Methods}

Recall that a class $\X$ is {\em meager\/} if it is included in a
countable union of nowhere dense subsets of $\Cant$, and {\em
comeager\/} if its complement $\cmp{\X}$ is meager.  The following
lemma (implicit in Rogers \cite[p341]{Roge67}) will be useful.

\begin{lemma} \label{le:category}
If $\X\in\bfSigma^0_2$ and $\cmp{\X}$ is dense then $\X$ is meager. 
\end{lemma}

\begin{proof}
Suppose that $\X=\bigcup_n \X_n$, $\X_n$ closed. 
Since $\cmp{\X}$ is dense, $\X_n$ contains no basic open set, 
hence $\X_n$ is nondense (i.e.\ its closure contains no 
basic open set), and $\X$ is a countable union of nondense sets. 
\end{proof}

\renewcommand{\DEC}{\Delta^0_1}
As a warm-up we give a short proof of Shoenfield's result that 
the class of computable sequences is not a $\Pi^0_3$-class.

\begin{theorem} {\rm (Shoenfield \cite[p344]{Roge67})}
The class $\DEC$ of computable sets is a $\Sigma^0_2[\Sigma^0_1]$-class, 
but it is not a $\Pi^0_3$-class.  It is also not a $\bmath{\Pi}^0_2$-class.
\end{theorem}
\begin{proof}
Clearly $\DEC\in \Sigma^0_2[\Sigma^0_1]$. Suppose for a contradiction 
that $\cmp{\DEC}$ is $\Sigma^0_3$. Then there is a uniform 
sequence of $\Sigma^0_1$-classes $\OO_{n,m}$ such that
$\cmp{\DEC}=\bigcup_n\bigcap_m\OO_{n,m}$. Without loss of 
generality $\OO_{n,m}\supseteq \OO_{n,m+1}$ for all $n$,$m$. 
Now $\DEC$ is meager because it is countable, so $\cmp{\DEC}$
is comeager, so there is an $n$ such that 
$\bigcap_m \OO_{n,m}$ is not nowhere dense, hence dense in 
some interval $C_\sigma$. Then every $\OO_{n,m}$, $m\in\N$, 
is dense in $C_\sigma$. Now it is easy to construct a computable 
sequence (starting with $\sigma$) in $\bigcap_m \OO_{n,m}$, contradicting
that $\bigcap_m \OO_{n,m}\subseteq\cmp{\DEC}$.

That $\DEC$ is not $\bmath{\Pi}^0_2$ follows from Lemma~\ref{le:category}, 
since $\cmp{\DEC}$ is comeager.
\end{proof}

The class $\RAND$ of Martin-L\"of random sets can 
easily be classified with category methods.  
\begin{theorem} {\rm (folk)} \label{th:RAND}
$\RAND$ is a $\Sigma^0_2$-class, but it is not a $\bfPi^0_2$-class.
\end{theorem}
\begin{proof}
This is analogous to the proof in Rogers \cite[p 341]{Roge67} that
$\{X: X \mbox{ finite}\}$ is a $\Sigma^0_2$-class but not a
$\Pi^0_2$-class.  Both $\RAND$ and its complement are dense, so by
Lemma~\ref{le:category}, $\RAND$ is meager. If $\RAND$ were a
$\bfPi^0_2$-class, then again using Lemma \ref{le:category}, its
complement would also be meager.  This contradicts the fact that
$\Cant$ is not meager.
\end{proof}
 
As $\DIM^0$ and $\DIMstr^1$ are dense $\bfPi^0_2$-classes that have
dense complements, an argument similar to the one used for Theorem
\ref{th:RAND} shows that they are not $\bfSigma^0_2$-classes.

\begin{theorem}\label{th:DIM_zero_DIMstr_one}
The classes $\DIM^0$ and $\DIMstr^1$ are not $\bfSigma^0_2$-classes.
\end{theorem}

We now develop category methods for the other $\DIM^\alpha$ classes.
For every rational $s$, define the computable order $h_s(n)=
2^{(1-s)n}$.  Let $\bd$ be the optimal constructive supermartingale.
\begin{lemma} \label{le:success_comeager}
For every rational $s \in (0,1)$, $S^{h_s}[\bd]$ is a comeager
$\Pi^0_2$-class.
\end{lemma}
\begin{proof}
Notice that $\cmp{S^{h_s}[\bd]}\in\Sigma^0_2$ and 
$S^{h_s}[\bd]$ is dense.  Now apply Lemma~\ref{le:category}. 
\end{proof}

\begin{lemma} \label{le:dim_meager}
For all $\alpha \in (0,1]$, $\DIM^\alpha$ is meager.
\end{lemma}
\begin{proof}
Let $s < \alpha$ be rational.  Lutz \cite{Lutz:DISS} showed that
$\bd^{(s)}(w)=2^{(s-1)|w|}\bd(w)$ is an optimal constructive
$s$-supergale.  It follows that for any $A \in \Cant$, $A \in
S^{h_s}[\bd] \Rightarrow \dim(S) < \alpha$. Therefore $\DIM^\alpha
\subseteq \cmp{S^{h_s}}$, so $\DIM^\alpha$ is meager by Lemma
\ref{le:success_comeager}.
\end{proof}

\begin{proposition}\label{prop:DIM}  
For all $\alpha \in (0,1]$, $\DIM^\alpha$ is not a $\Pi^0_2$-class.
\end{proposition}
\begin{proof}
If $\DIM^\alpha \in \Pi^0_2$, then Lemma \ref{le:category} implies that
$\DIM^\alpha$ is comeager, contradicting Lemma \ref{le:dim_meager}.
\end{proof}

To strengthen Proposition \ref{prop:DIM} to show that $\DIM^\alpha$ is
not $\Sigma^0_3$, we now turn to Wadge reductions.

\subsection{Wadge Reductions}

Let $\A,\B \subseteq \Cant$.  A {\em Wadge reduction} of $\A$ to $\B$
is a function $f : \Cant \to \Cant$ that is continuous and satisfies
$\A = f^{-1}(\B)$, i.e., $X \in \A \iff f(X) \in \B$.  We say that
$\B$ is {\em Wadge complete} for a class $\bmath{\Gamma}$ 
of subsets of $\Cant$
if $\B \in \bmath{\Gamma}$ and every 
$\A \in \bmath{\Gamma}$ Wadge reduces to $\B$.  As
the classes of the Borel hierarchy are closed under Wadge reductions,
Wadge completeness can be used to properly identify the location of a
subset of $\Cant$ in the hierarchy.

We now prove that $\DIM^1$ is Wadge complete for $\bfPi^0_3$.  We
will then give Wadge reductions from it to $\DIM^\alpha$ for the other
values of $\alpha$.

\begin{theorem} \label{th:DIM_one}
$\DIM^1$ is Wadge complete for $\bfPi^0_3$.  Therefore $\DIM^1$ is
  not a $\bfSigma^0_3$-class, and in particular it is not a $\Sigma^0_3$-class.
\end{theorem}
\begin{proof}
One could prove this by reducing a known $\bfPi^0_3$-complete
class to $\DIM^1$, e.g.\ the class of sets that have a
limiting frequency of 1's that is 0 (this class was proved 
to be $\bfPi^0_3$-complete by Ki and Linton \cite{KiLinton94}), 
but it is just as easy to build a direct reduction from an 
arbitrary $\bfPi^0_3$-class. 

Let $\bd$ be the universal constructive supermartingale.  Note that  
we have (cf.\ Proposition~\ref{equiv}) 
$$
S^{2^n}[\bd] \subsetneq \ldots
\subsetneq S^{2^{\frac{1}{k}n}}[\bd] \subsetneq
S^{2^{\frac{1}{k+1}n}}[\bd] \subsetneq \ldots
\subsetneq \DIM^1.
$$
Let $\bigcup_k\bigcap_s \OO_{k,s}$ be a $\bfSigma^0_3$-class. 
Without loss of generality $\OO_{k,s}\supseteq \OO_{k,s+1}$ for all $k$,$s$.
We define a continuous function $f:\Cant\rightarrow \Cant$ 
such that 
\begin{equation}
\forall k \left( X\in\bigcap_s \OO_{k,s} \Longleftrightarrow
f(X) \in S^{2^{\frac{1}{k}n}}[\bd] \right)
\end{equation}
so that we have 
\begin{eqnarray*}
X\not\in \bigcup_k\bigcap_s \OO_{k,s}  & \Longleftrightarrow  & 
\forall k \left(  f(X)\not\in S^{2^{\frac{1}{k}n}}[\bd] \right) \\
                                       & \Longleftrightarrow  & f(X)\in\DIM^1.
\end{eqnarray*}
The image $Y=f(X)$ is defined in stages, $Y=\bigcup_s Y_s$, 
such that every initial segment 
of $X$ defines an initial segment of $Y$. 

At stage 0 we define $Y_0$ to be the empty sequence.

At stage $s>0$ we consider $X\restr s$, and for each $k$ we
define $t_{k,s}$ to be the largest stage $t \leq s$ such that 
$X\restr s \in \OO_{k,t}$.  (Let $t_{k,s} = 0$ if such a $t$ does not exist.) 
Define $k$ to be {\em expansionary} at stage $s$ if $t_{k,s-1} < t_{k,s}$.
Now we let $k(s) = \min \{k : k \textrm{ is expansionary at }s \}$.
There are two substages. 

{\em Substage (a).}  First consider all strings $\sigma$ extending
$Y_{s-1}$ of minimal length with $\bd(\sigma)\geq
2^{\frac{1}{k(s)}|\sigma|}$, and take the leftmost one of these
$\sigma$'s.  Such $\sigma$'s exist because
$S^{2^{\frac{1}{k(s)}n}}[\bd]$ is dense.  
If $k(s)$ does not exist, let
$\sigma=Y_{s-1}$.

{\em Substage (b).}  Next consider all extensions
$\tau\sqsupseteq\sigma$ of minimal length such that $\bd(\tau\restr
i)\leq \bd(\tau\restr(i-1))$ for every $|\sigma|<i<|\tau|$, and
$\bd(\tau)\leq |\tau|$.  Clearly such $\tau$ exist, by direct
diagonalization against $\bd$.  Define $Y_s$ to be the leftmost of
these $\tau$.  This concludes the construction.

So $Y_s$ is defined by first building a piece of evidence 
$\sigma$ that $\bd$ achieves growth rate $2^{\frac{1}{k(s)}n}$ on $Y$ and 
then slowing down the growth rate of $\bd$  to the order $n$.
Note that $f$ is continuous.
If $X\in\bigcup_k\bigcap_s \OO_{k,s}$, then for the minimal 
$k$ such that $X\in \bigcap_s \OO_{k,s}$, infinitely many 
pieces of 
evidence $\sigma$ witness that $\bd$ achieves growth rate 
$2^{\frac{1}{k}n}$ on $Y$, so $Y\not\in\DIM^1$.
On the other hand, if $X\not\in\bigcup_k\bigcap_s \OO_{k,s}$
then for every $k$ only finitely often $\bd(Y_s)\geq 2^{\frac{1}{k}|Y_s|}$
because in substage (a) the extension $\sigma$ is chosen to be of 
minimal length, 
so $Y\not\in S_{h_k}[\bd]$. Hence $Y\in\DIM^1$. 
\end{proof}

As $\RAND$ is a $\Sigma^0_2$-class, we have the following corollary
(which can also be proved by a direct construction).

\begin{corollary_cite}{Lutz \cite{Lutz:DISS}} \label{proper}
$\RAND$ is a proper subset of $\DIM^1$. 
\end{corollary_cite}

In order to establish the existence of $\Delta^0_2$-computable
sequences of any $\Delta^0_2$-computable dimension $\alpha \in [0,1)$,
Lutz \cite{Lutz:DISS} defined a {\em dilution function} $g_\alpha : \Cant
\to \Cant$ that is computable and satisfies $\dim(g_\alpha(X)) =
\alpha \cdot \dim(X)$ for all $X \in \Cant$.  Applying this to any
$\Delta^0_2$-computable Martin-L\"of random sequence (which must have
dimension 1) establishes the existence theorem.  
(We note that $g_\alpha(X)$ has the same Turing degree as $X$. 
Since by the Low Basis Theorem of Jockusch and Soare \cite[Theorem V.5.32]{Odif89}
there are 
Martin-L\"of random sets of low degree, we immediately obtain 
that there are low sets of any $\Delta^0_2$-computable dimension $\alpha$.)
As $g_\alpha$ is
continuous, it is a Wadge reduction from $\DIM^1$ to $\DIM^\alpha$ if
$\alpha > 0$.  Combining this with the previous theorem, we have
that $\DIM^\alpha$ is Wadge complete for $\bfPi^0_3$ for all
$\Delta^0_2$-computable $\alpha \in (0,1)$.  We now give a similar
dilution construction that will allow us to prove this for {\em
arbitrary} $\alpha \in (0,1)$.

Let $X \in \Cant$ and let $\alpha \in (0,1)$.  Write $X = x_1 x_2 x_3
\ldots$ where $|x_n| = 2n-1$ for all $n$, noting that $|x_1\cdots x_n|
= n^2$.  For each $n$, let
$$k_n = \ceil{ n \frac{1-\alpha}{\alpha} }$$ and $y_n = 0^{k_n}$.  We
then define
$$f_{\alpha}(X) = x_1 y_1 x_2 y_2 \cdots x_n y_n \cdots\ .$$
Observe that $f_\alpha$ is a continuous function mapping $\Cant$ to $\Cant$.
We now show that it modifies the dimension of $X$ in a controlled
manner.

\begin{lemma}\label{le:dilution}  For any $X \in \Cant$ and $\alpha
  \in (0,1)$,
$$\dim(f_\alpha(X)) = \alpha \cdot \dim(X)$$
and
$$\Dim(f_\alpha(X)) = \alpha \cdot \Dim(X).$$
\end{lemma}

\begin{proof}
The proof uses \eqref{eq:K_dim} and \eqref{eq:K_Dim}, the Kolmogorov complexity
characterizations of dimension and strong dimension.

Let $w \prefix f_\alpha(X)$.  For some $n$,
$$w = x_1 y_1 \cdots x_{n-1} y_{n-1} v,$$
where $v \prefix x_n y_n$.  Then
\begin{eqnarray*}
K(w) 
&\leq& K(x_1 \cdots x_{n-1}) + K(v) \\
&& + K(k_1) + \cdots + K(k_{n-1}) + O(1) \\
&\leq& K(x_1 \cdots x_{n-1}) + O(n \log n).
\end{eqnarray*}
Because
$$|w| \geq |x_1y_1\cdots x_{n-1}y_{n-1}| \geq
\frac{(n-1)^2}{\alpha},$$
we have
$$\frac{K(w)}{|w|} \leq \frac{\alpha\cdot K(x_1\cdots x_{n-1})}{|x_1\cdots
  x_{n-1}|} + \frac{O(n\log n)}{(n-1)^2},$$
It follows that
\begin{eqnarray*}
\dim(f_\alpha(X)) &\leq& \alpha \liminfn \frac{K(x_1\cdots
  x_{n-1})}{|x_1\cdots x_{n-1}|} \\
&=& \alpha \liminfn \frac{K(x \restr n)}{n} \\
&=& \alpha \cdot \dim(X),
\end{eqnarray*}
where the first equality holds because the block $x_n$ is short
relative to $x_1 \cdots x_{n-1}$.  Similarly, $\Dim(f_\alpha(X)) \leq
\alpha \cdot \Dim(X)$.

For the other inequality, we have
\begin{eqnarray*}
K(x_1 \cdots x_{n-1}) &\leq& K(w) + K(k_1) + \cdots + K(k_{n-1}) \\&&+
O(1)\\
&\leq& K(w) + O(n\log n)
\end{eqnarray*}
and
$$|w| \leq |x_1 y_1 \cdots x_n y_n| \leq \frac{n^2}{\alpha} + n \leq \frac{(n+1)^2}{\alpha},$$
so
\begin{eqnarray*}
\frac{K(w)}{|w|} &\geq& \alpha \frac{K(x_1 \cdots x_{n-1}) - O(n\log n)}{(n+1)^2}\\
&=& \alpha \frac{K(x_1 \cdots x_{n-1})}{|x_1\cdots x_{n-1}|}
\frac{(n-1)^2}{(n+1)^2} - \frac{O(n\log n)}{(n+1)^2}.
\end{eqnarray*}
Therefore
\begin{eqnarray*}
\dim(f_\alpha(X)) &\geq& \alpha \liminfn \frac{K(x_1\cdots
  x_{n-1})}{|x_1\cdots x_{n-1}|} \\
&=& \alpha \liminfn \frac{K(x \restr n)}{n} \\
&=& \alpha \cdot \dim(X),
\end{eqnarray*}
and analogously, $\Dim(f_\alpha(X)) \geq \alpha \cdot \Dim(X)$.
\end{proof}
 
The function $f_\alpha$ establishes the completeness of 
$\DIM^\alpha$.

\begin{theorem}  For all $\alpha \in (0,1)$, $\DIM^\alpha$
is Wadge complete for $\bfPi^0_3$.  Therefore it is not a
$\bfSigma^0_3$-class, and in particular it is not a $\Sigma^0_3$-class.
\end{theorem}
\begin{proof}
By Lemma \ref{le:dilution}, $f_\alpha$ is a Wadge reduction from
$\DIM^1$ to $\DIM^\alpha$.  Therefore $\DIM^\alpha$ is Wadge complete
for $\bfPi^0_3$ by composing $f_\alpha$ with the reduction from
Theorem \ref{th:DIM_one}.
\end{proof}

As $g_\alpha$ is also a Wadge reduction from $\DIMstr^1$ to
$\DIMstr^\alpha$, we have from Theorem \ref{th:DIM_zero_DIMstr_one}
that $\DIMstr^\alpha$ is not a $\bfSigma^0_2$-class for all $\alpha
\in (0,1)$.  We now prove that $\DIMstr^\alpha$ is not even
$\bfSigma^0_3$ for all $\alpha \in [0,1)$.

\begin{theorem} \label{th:DIMstr_zero}
For all $\alpha \in [0,1)$, $\DIMstr^\alpha$ is Wadge complete for
  $\bfPi^0_3$.  Therefore $\DIMstr^\alpha$ is not a
  $\bfSigma^0_3$-class, and in particular it is not a
  $\Sigma^0_3[\Pi^0_1]$-class.
\end{theorem}

\begin{proof}
The proof is similar to that of Theorem \ref{th:DIM_one}, but uses
\eqref{eq:K_Dim}, the Kolmogorov complexity characterization of strong
dimension.  Let $\C = \bigcup_k\bigcap_s \OO_{k,s}$ be a
$\bfSigma^0_3$-class and without loss of generality assume that
$\OO_{k,s}\supseteq \OO_{k,s+1}$ for all $k$,$s$.

Let $\alpha \in (0,1)$.  (We will discuss the simpler case $\alpha =
0$ later.)  We define a continuous function $f:\Cant\rightarrow \Cant$
in stages that will Wadge reduce $\C$ to $\cmp{\DIMstr^\alpha}$.  The
image $Y = f(X)$ will be the unique sequence extending $Y_s$ for all
$s$.  At stage 0 we define $Y_0$ to be the empty sequence.

At stage $s>0$ we consider $X\restr s$, and define $k(s)$ as in the
proof of Theorem \ref{th:DIM_one}.  There are three substages. 

{\em Substage (a).}  First consider all strings $\rho$ extending
$Y_{s-1}$ of minimal length with $K(\rho) \geq \alpha |\rho|$, and
take the leftmost one of these $\rho$'s.

{\em Substage (b).}  Next consider all strings $\sigma$ extending
$\rho$ of minimal length with $K(\sigma) \geq (\alpha +
\frac{1}{k(s)})|\sigma|$, and take the leftmost one of these
$\sigma$'s.  If $k(s)$ does not exist, let $\sigma=\rho$.

{\em Substage (c).}  Extend $\sigma$ with a block of 0's to obtain
$Y_s = \sigma 0^{|\sigma|^2-|\sigma|}$.

That is, to define $Y_s$, we first select $\rho$ to increase the
Kolmogorov complexity rate to $\alpha$.  This ensures that $Y$ will
have strong dimension at least $\alpha$.  We then construct a piece of
evidence $\sigma$ that $Y$ has strong dimension at least $\alpha +
\frac{1}{k(s)}$.  We finish $Y_s$ with a long block of 0's to bring
the Kolmogorov complexity down to a near-zero rate, so that the next
stage will work properly.

If $X\in\C$, then for the minimal $k$ such
that $X\in \bigcap_s \OO_{k,s}$, infinitely many prefixes $\sigma
\prefix Y$ satisfy $K(\sigma) \geq (\alpha + \frac{1}{k})|\sigma|$.
Therefore 
$\Dim(Y) \geq \alpha + \frac{1}{k}$, so $Y \not\in \DIMstr^\alpha$.

Now let $X\not\in \C$.  Let $\alpha' > \alpha$ be arbitrary, and
choose $k$ so that $\frac{1}{k} < \alpha' - \alpha$.  Because $X
\not\in \C$, we have $k(s) > k$ for all sufficiently large $s$.
Let $s_0$ be large enough to ensure $k(s) > s$ and $K(Y_{s-1}) \leq
\sqrt{|Y_{s-1}|} + O(1) < \alpha |Y_{s-1}|$ hold for all $s \geq s_0$.
Suppose that 
\begin{equation}\label{eq:high_K}
K(w) \geq \alpha' |w|.
\end{equation}
holds for some $w$ with $Y_{s-1} \prefix w \prefix Y_{s}$ for some
stage $s \geq s_0$.
We then
have that $\rho$ is a proper extension of $Y_{s-1}$.  By choice of
$\rho$ and $\sigma$ and the fact that $\alpha' > \alpha + \frac{1}{k}
> \alpha + \frac{1}{k(s)}$, we must have $w = \rho$ or $\sigma \prefix
w$.  We analyze these two cases separately.
\begin{enumerate}[\upshape (i)]
\item $w = \rho$: Let $\rho'$ be the string obtained from
$\rho$ by removing the last bit.  Then $K(\rho) \leq K(\rho') + O(1)$.
By choice of $\rho$, we have $K(\rho') < \alpha |\rho'|$.  We also
have $K(\rho) \geq (\alpha') |\rho|$ by \eqref{eq:high_K}.
Putting these three statements together yields
$$\alpha' |\rho| < \alpha (|\rho|-1) + O(1),$$
which is a contradiction if $|\rho| = |w|$ is sufficiently large.
\item $\sigma \prefix w$: Obtain $\sigma'$ from $\sigma$ by
removing the last bit of $\sigma$.  Then we have
\begin{eqnarray*}
K(w) &\leq& K(\sigma') + K(|w|-|\sigma|) + O(1)\\
& \leq & K(\sigma') + \log (|w|-|\sigma|) + O(1) \\ 
&\leq& K(\sigma') + 2 \log |\sigma| + O(1).
\end{eqnarray*}
By choice of $\sigma$, $K(\sigma') < (\alpha +
\frac{1}{k(s)})|\sigma'|$.  These two facts together with
\eqref{eq:high_K} tell us that
$$\alpha'|w| < \left(\alpha + \frac{1}{k(s)}\right)(|\sigma|-1) +
2\log |\sigma| + O(1),$$ which is a contradiction for large $|w|$
because $|w| \geq |\sigma|$ and $\alpha' >
\alpha + \frac{1}{k(s)}$.
\end{enumerate}
Therefore, for all sufficiently long $w \prefix Y$,
\eqref{eq:high_K} does not hold.  It follows that $\Dim(Y) \leq
\alpha$.  On the other hand, there are infinitely many $\rho \prefix
Y$ with $K(\rho) \geq \alpha |\rho|$, so $\Dim(Y) \geq \alpha$.
Therefore $Y \in \DIMstr^\alpha$.

This shows that $f$ is a Wadge reduction from $\C$ to
$\cmp{\DIMstr^\alpha}$.  As $\C$ is an arbitrary
$\bfSigma^0_3$-class, this shows that $\DIMstr^\alpha$ is Wadge
complete for $\bfPi^0_3$.

The proof for the case $\alpha = 0$ is similar, but simpler as
substage (a) is omitted in the construction.
\end{proof}

\subsection{Ad Hoc Methods} \label{adhoc}

When classifying classes in the arithmetical hierarchy of reals 
there are several methods one can use. As we have seen, 
category methods are sometimes useful up to the second level,
Wadge reductions are useful if the classification in the effective (lightface) 
hierarchy coincides with that in the classical (boldface) hierarchy, and sometimes
(as in Proposition~\ref{prop:borel} and Example~\ref{generic}) 
one just needs something else.
In particular when the level of the class in the effective hierarchy is 
not the same as the level in the classical hierarchy one often needs to resort 
to ad hoc arguments. 
One might think that the notion of effective Wadge reduction, or 
recursive functional, would  be the proper notion to use in classifying 
classes of reals in the effective hierarchy. 
However, this notion is rarely useful for the following reason. 
Let $\X$ be a class without computable elements, such as the class of 
Martin-L\"of random sets or the class of 1-generic sets. 
Then $\X$ cannot be proven to be complete for any level of the 
effective hierarchy by a recursive Wadge reduction $f$.
For if $X$ is recursive, then so is $f(X)$, so we can never have 
$X \in \Cant \Longleftrightarrow f(X) \in \X$. So we see that 
``easy'' classes like $\Cant$ that contain recursive elements cannot 
be reduced in such a way to many ``difficult'' classes, which 
renders the notion rather useless. 

We have left open the question whether $\DIMstr^1$ is not in $\Pi^0_2$, 
and whether $\DIMstr^\alpha$ is not in $\Pi^0_3$
for any $\Delta^0_2$-computable $\alpha \in [0,1)$. 
We have no answer to the second question, but we provide an 
answer to the first in the next theorem. We make use of the following lemma.

\begin{lemma} \label{e-z-as-pi}
If $\X \in\Pi^0_2$ is dense then there is a computable $X\in\X$. 
\end{lemma}
\begin{proof}
This is an easy finite extension argument.
Suppose that $\X = \{X : (\forall m)(\exists k) R^X(m,k)\!\downarrow = 1 \} \in \Pi^0_2$ is dense. 
(Here $R$ is a computable predicate. Note that $R$ does not have to be defined 
with oracles $X$ that are not in $\X$.) 
Given any initial segment $\tau$ such that 
$$
(\forall n < m)(\exists k) R^\tau(m,k)\!\downarrow = 1,
$$
we show how to compute an extension $\sigma \sqsupset \tau$ such that 
\begin{equation} \label{wunderbar}
(\exists k) R^\sigma(m,k)\!\downarrow = 1.
\end{equation}
Because $\X$ is dense, there are $X\sqsupset\tau$ and $k$ such that 
$R^X(m,k)\!\downarrow = 1$. Let $u$ be the use of this computation, i.e.\
the part of the oracle $X$ used in it. Now define $\sigma = \max \{X\!\restr\! u, \tau\}$.
Then $\sigma\sqsupseteq\tau$ satisfies (\ref{wunderbar}).

Now it is clear that for every $m$ we can compute appropriate extensions $\sigma_m$ 
such that $X=\bigcup_{m} \sigma_m$ is computable and 
$(\forall m)(\exists k) R^{\sigma_m}(m,k)\!\downarrow = 1$, so that $X\in\X$. 
\end{proof}

\begin{theorem}
$\DIMstr^1$ is not a $\Pi^0_2$-class. Hence it is properly $\Pi^0_2[\Pi^0_1]$.
\end{theorem}
\begin{proof}
Suppose that $\DIMstr^1$ is $\Pi^0_2$. Then, since clearly $\DIMstr^1$ is dense, 
by Lemma~\ref{e-z-as-pi} it contains a computable real, contradicting 
that every computable real has strong dimension~0. 
\end{proof}

We conclude this section by summarizing its main results in the following
table.

\begin{center}
\begin{tabular}{|c|c|c|c|}
\hline
 & $\DIM^\alpha$ & $\DIMstr^\alpha$ \\ \hline\hline
$\alpha = 0$ & $\Pi^0_2 - {\bfSigma^0_2}$ & $\Pi^0_3[\Sigma^0_1] -
 {\bfSigma^0_3}$ \\ \hline
$\alpha \in (0,1)\cap\Delta^0_2$ & $\Pi^0_3-{\bfSigma^0_3}$ & $\Pi^0_3[\Sigma^0_1] -
 {\bfSigma^0_3}$ \\ \hline
$\alpha = 1$ & $\Pi^0_3-{\bfSigma^0_3}$ & $\Pi^0_2[\Pi^0_1] -
 ({\bfSigma^0_2} \cup \Pi^0_2)$ \\ \hline
arbitrary $\alpha \in (0,1)$ & ${\bfPi^0_3}-{\bfSigma^0_3}$ & ${\bfPi^0_3} -
 {\bfSigma^0_3}$ \\ 
\hline
\end{tabular}
\end{center}

\begin{question}
Is it the case that $\DIMstr^\alpha$ is not in $\Pi^0_3$  
for any $\Delta^0_2$-computable $\alpha \in [0,1)$? 
\end{question}

\section{Effective Randomness Classes} \label{sec:randclass}

We begin this section by pointing out some relationships between
computable dimension, Church randomness, and Schnorr randomness.

Analogously to what was done for the constructive case, the {\em
computable dimension} of a class $\A\subseteq \Cant$ is defined as
$$ \dimcomp(\A) = 
\inf \left\{ s \left| \begin{array}{l}\textrm{there
exists a computable}\\\textrm{$s$-gale $d$ for which $\A
\subseteq S^\infty[d]$}\end{array} \right.\right\}.$$

A {\em selection rule\/} is a function
$\varphi:\{0,1\}^*\rightarrow\{0,1\}$.
With every selection rule $\varphi$ we associate a function
$\Phi:\{0,1\}^*\rightarrow\{0,1\}^*$ defined by
$\Phi(\lambda)=\lambda$ and
$$
\Phi(wi) = \left\{\begin{array}{ll}
\Phi(w)i & \mbox{if $\varphi(w)=1$,}\\
\Phi(w)  & \mbox{if $\varphi(w)=0$.}
\end{array}\right.
$$ A set $A$ is called {\em Church random\/} if every substring of
$\chi_A$ (the characteristic string of $A$) defined by a computable
selection rule is stochastic, i.e., satisfies the law of large
numbers.  Consider the following property of selection rules:
\begin{equation} \label{selection}
\inf_{w\in\{0,1\}^*} \frac{|\Phi(w)|}{|w|} >0.
\end{equation}

A {\em computable null set of exponential order\/} is a set of the
form $S^{a^n}[d]$, where $d$ is a computable martingale and $a > 1$.
It is easy to check that a set is not in any computable null set of
exponential order if and only if $\{A\}$ has computable dimension~1.
With this observation, we can restate a result of Schnorr as follows.

\begin{theorem} {\rm (Schnorr \cite[Satz 17.8]{Schn71b})} \label{exponential}
$\{A\}$ has computable dimension~1 if and only if every substring of
$\chi_A$ selected by a computable selection rule with the property
(\ref{selection}) is stochastic.
\end{theorem}
In particular every Church random sequence is not in any null set of
the form $S^{a^n}[d]$ where $d$ is computable. In the words of Schnorr
\cite{Schn77}, ``Church random sequences approximate the behavior of
Schnorr random sequences.''

\begin{proposition}\label{prop:Church}
There are sequences with computable dimension 1 that are not Church
random.
\end{proposition}

\begin{proof}
Let $R$ be computably random, and let $D=\{2^n \mid n \in\N\}$ be an
exponentially sparse decidable domain.  Then $A=R-D$ has computable
dimension~1, but $D$ can be computably selected, so $A$ is
not Church random.
\end{proof}

We now classify the Schnorr random sequences in the arithmetical
hierarchy.

\begin{theorem} \label{th:Schnorr_rand}
$\RANDSch$ is a $\Pi^0_3$-class, but not a $\bfSigma^0_3$-class.
\end{theorem}

\begin{proof}
First note that $\RANDSch\in\Pi^0_3$: $X\in\RANDSch$ if and only if for
every pair of codes $e$ and $f$, either the $e$-th partial computable
function $\varphi_e$ is not a computable order (i.e.\ is not total or
decreases at some point), $\varphi_f$ is not a computable martingale
(i.e.\ is not total or violates the martingale property at some
point), or $X\not\in S^{\varphi_e}[\varphi_f]$, and that every one of
these options is $\Sigma^0_2$.

The rest of the proof resembles that of Theorem~\ref{th:DIM_one}.  Fix
a (non-computable) sequence of computable martingales
$\{d_k\}_{k\in\N}$ and a sequence of computable orders
$\{h_k\}_{k\in\N}$ such that
\begin{enumerate}[\rm (i)]

\item $X\in\RANDSch$ $\Longleftrightarrow \forall k (X\not\in
S^{h_k}[d_k])$.

\item $S^{h_k}[d_k] - S^{\min\{ h_j:j<k\}}[\sum_{j<k} d_j]$ 
is dense for every $k$. 

\end{enumerate}
The $d_k$ can be defined by taking appropriate sums of computable
martingales so that for any computable martingale $d$, there is some
$d_k$ such that $d_k(w) \geq d(w)$ for all $w$.  For the $h_k$ one can
take any family of computable orders such that every computable order
$h$ dominates some $h_k$. (Of course the $d_k$ and $h_k$ cannot be
uniformly computable families, but that is of no concern to us.)

Let $\bigcup_k\bigcap_s \OO_{k,s}$ be a $\bfSigma^0_3$-class.
We define a continuous function $f:\Cant\rightarrow \Cant$
such that
\begin{equation}
\forall k \left( X\in\bigcap_s \OO_{k,s} \Longleftrightarrow
f(X) \in S^{h_k}[d_k] \right)
\end{equation}
so that by (i) we have 
$X\not\in \bigcup_k\bigcap_s \OO_{k,s}$ $\Longleftrightarrow$
$f(X)\in\RANDSch$.

As in the proof of Theorem~\ref{th:DIM_one} we define the 
image $Y=f(X)$ in stages. Every time we find a new piece of evidence 
that $X\in\bigcap_s \OO_{k,s}$, at stage $s$ say, 
 we build a piece of evidence that 
$Y\in S^{h_k}[d_k]$ by choosing an appropriate finite extension at 
stage $s$. Such an extension can be found by (ii). 
The rest of the proof is identical to that of Theorem~\ref{th:DIM_one}.
\end{proof}

With only some obvious changes one can also prove the following theorem.

\begin{theorem} \label{th:rec_rand}
$\RANDcomp$ is a $\Pi^0_3$-class, but not a $\bfSigma^0_3$-class.
\end{theorem}

\begin{proof}
Note that $X$ is computably random if and only if for every $e$,
$\varphi_e$ is not a computable martingale or $X\not\in \regSS[\varphi_e]$,
so the class is $\Pi^0_3$.  That it is properly $\Pi^0_3$ is actually
easier than the proof of Theorem~\ref{th:Schnorr_rand} since we only
need the sequence $\{d_k\}$ and not the $\{h_k\}$.  \end{proof}

In contrast to the universal constructive supermartingale $\bd$
satisfying $\RAND = \cmp{\regSS[\bd]}$, Theorem \ref{th:rec_rand}
implies that, even from a noncomputable standpoint, $\RANDcomp$ has no
such universal object.  That is, $\RANDcomp \not= \cmp{\regSS[d]}$ for
any (arbitrarily noncomputable) supermartingale $d$, as otherwise
$\RANDcomp$ would be a $\bfSigma^0_2$-class.

In this paper we have considered only the extension of the 
artimetical hierarchy of reals by adding one local quantifier. 
We end by remarking that one can add of course 
more local quantifiers. The classes thus obtained also 
have natural inhabitants. To give an example, again 
from the theory of randomness, 
recall that a set $A$ is $n$-random if it is 
Martin-L\"of random relative to $\emptyset^{(n-1)}$.
So it is 1-random if it is Martin-L\"of random,
2-random if it is Martin-L\"of random relative to $K$, etc.
Now the class of $n$-random sets is $\bfSigma^0_2$ for 
every $n$, and in fact one can check that it is 
$\Sigma^0_2[\Sigma^0_{n-1}]$.

\begin{ack}The second and third authors thank Alekos Kechris for a
helpful discussion.
\end{ack}


\end{document}